\documentclass{cjpsuppl} 
\usepackage{cite,graphics}
\newcommand{\Dmb}[1][]{\ensuremath{\Delta m_b^{#1}}}

\newcommand{\Dmf}[1][]{\ensuremath{\Delta m_f^{#1}}}
\newcommand{\tb}[1][]{\ensuremath{\tan^{#1}\!\beta}}
\newcommand{\Dmtau}{\ensuremath{\Delta m_\tau}}
\newcommand{\fb}{\ensuremath{\rm \,fb}}
\newcommand{\pb}{\ensuremath{\rm \,pb}}
\newcommand{\mHc}{\ensuremath{m_{H^\pm}}}

\begin{document} 
\title{Distinguishing Higgs models in 
$H^+\rightarrow \tau^+\nu /  t\bar{b}$ at large $\tan
\beta$~\footnote{Based 
on the poster presented by S.~Pe{\~n}aranda at {\em Physics
    at LHC}, 13-17 July 2004, Vienna, Austria.
Preprint numbers: CERN-PH-TH/2004-170, SHEP-04-26.}}
\authori{K.A.~Assamagan}
\addressi{Brookhaven National Laboratory, Upton NY 11792, USA} 
\authorii{J.~Guasch}
\addressii{Paul Scherrer Institut, CH-5232 Villigen PSI,
  Switzerland}
\authoriii{S.~Moretti}
\addressiii{School of Physics and Astronomy,
Southampton University, Highfield SO17 1BJ, UK}
\authoriv{S.~Pe{\~n}aranda$^{\,*}$}
\addressiv{CERN-TH Division, Department of Physics, CH-1211
  Geneva 23, Switzerland} 
\authorv{}     \addressv{}
\authorvi{}    \addressvi{}
\headtitle{Distinguishing Higgs models in 
$H^+\to \tau^+\nu / H^+\to t\bar{b}$ at large $\tan \beta$}
\headauthor{K.A.~Assamagan et al.}
\lastevenhead{K.A.~Assamagan et al.: Distinguishing Higgs models in 
$H^+\to \tau^+\nu / H^+\to t\bar{b}$ at large $\tan \beta$}
\refnum{}
\daterec{}
\suppl{A}  \year{2004} 
\maketitle

\begin{abstract}
We present an experimental and theoretical analysis of the ratio of
branching ratios $R=BR(H^+\to \tau^+\nu)/BR(H^+\to t\bar{b})$
of charged Higgs boson decays as a discriminant quantity
between supersymmetric and non-supersymmetric models. 
\end{abstract}
 
The detection of a charged Higgs boson $H^{\pm}$ would be a distinctive
signal of physics beyond the Standard Model (SM), since such a particle
does not exist in the SM. 
The associated production of a charged Higgs boson with a top quark
  ($pp\to H^+ \bar{t}+X$)~{\cite{Gunion:1994sv}} is 
  mainly relevant at large values of $\tan \beta$, a regime where
  Higgs boson observables receive large Supersymmetric (SUSY) radiative
  corrections. Here we report on the investigation of 
the production of charged Higgs bosons in
association with top quarks at the Large Hadron Collider (LHC), 
from the experimental and
theoretical points of view, by studying hadronic ($H^+\to t\bar{b}$) and
leptonic ($H^+\to \tau^+\nu$) decay signatures. Details of the analysis
have been presented in~\cite{our}.

While SUSY radiative effects might be difficult to discern in the
  production cross-sections separately, they will appear neatly in the 
relation:
  \begin{equation}
    \label{eq:relation}
    \label{eq:relation2}
    R\equiv\frac{\sigma(pp\to H^+\bar{t} + X \to \tau^+ \nu t + X)}{\sigma(pp\to H^+\bar{t} + X \to t\bar{b} \bar{t} +
    X)}=\frac{BR(H^+\to \tau^+\nu)}{BR(H^+\to t\bar{b})}=\frac{\Gamma(H^+\to \tau^+\nu)}{\Gamma(H^+\to t\bar{b})} \,\,.
  \end{equation}
As eq.~(\ref{eq:relation}) shows, the dependence on the
production mode (and on its large sources of
  uncertainty deriving from parton luminosity, unknown QCD radiative
  corrections, scale choices, etc.) cancels out.

On the other hand, the couplings of the Higgs particles to down-type
fermions receive large quantum corrections in the Minimal Supersymmetric
Standard Model (MSSM), enhanced by $\tan \beta=v_2/v_1$. These
corrections
have been resummed to all orders in perturbation theory, with the help of
the effective Lagrangian formalism  for the $t\bar{b}H^+$
vertex~\cite{Carena:1999py}. The relation between 
the fermion mass and the Yukawa coupling is modified by 
quantum corrections, encoded in a quantity $\Dmf$ ($f=b,\tau$), which is
non-decoupling  and contains all (potentially) large leading radiative
effects. The ratio~(\ref{eq:relation2}) receives large one-loop MSSM 
radiative corrections at large 
$\tan \beta$, due to $\Dmb$ and $\Dmtau$, which we have considered in
the present analysis.

We have performed a detailed phenomenological analysis 
of charged Higgs boson signatures for the LHC, by using the two-body production
cross-section subprocess $g\bar{b}\to H^+\bar{t}$, with the hadronic
 and leptonic decay channels  
of the charged Higgs boson. We have normalized our
production cross-section to the LO {result}, for consistency with the
tree-level treatment of the backgrounds.
In our simulation, we have let the top quarks decay through the SM-like channel
$t\to W^+b$. The
experimental signatures of the two production channels under investigation are
($l=e,\mu$):
\begin{eqnarray}
\label{eq:leptonic} 
  pp(g\bar{b})&\to & H^+ \bar{t} \to (\tau^+\nu) \bar{t} \to \tau^+\nu\,
  (jj \bar{b})\,\,,\\
  \label{eq:hadronic}
  pp(g\bar{b})&\to & H^+ \bar{t} \to (t\bar{b}) \bar{t} \to (jj [l\nu]
  b)\, \bar{b} \, (l\nu [jj]\bar{b})\,\,. 
\end{eqnarray}
Further details of the studies presented here are
available in~\cite{our,ketevi}.

The Monte Carlo (MC) simulation has been performed using  {\tt PYTHIA}
({v6.217})~\cite{Sjostrand:2000wi} for the signal and most of the background
processes. We have cross-checked the signal cross-section
with results presented in~\cite{Plehn:2002vy}. 
Discrepancies have been resolved by fixing a
bug in {\tt PYTHIA}({v6.217} or older). We have used {\tt HDECAY}
for the Higgs boson decay rates, {\tt TAUOLA} for the
$\tau$-lepton decays, {\tt TopRex}  for some 
background processes with a custom 
interface to {\tt PYTHIA}, and {\tt ATLFAST} for the 
detector simulation~\cite{programs}. 

\begin{figure}[t]
\begin{center}\vspace*{-0.5cm}
\begin{tabular}{cc}
\resizebox{6.3cm}{!}{\includegraphics{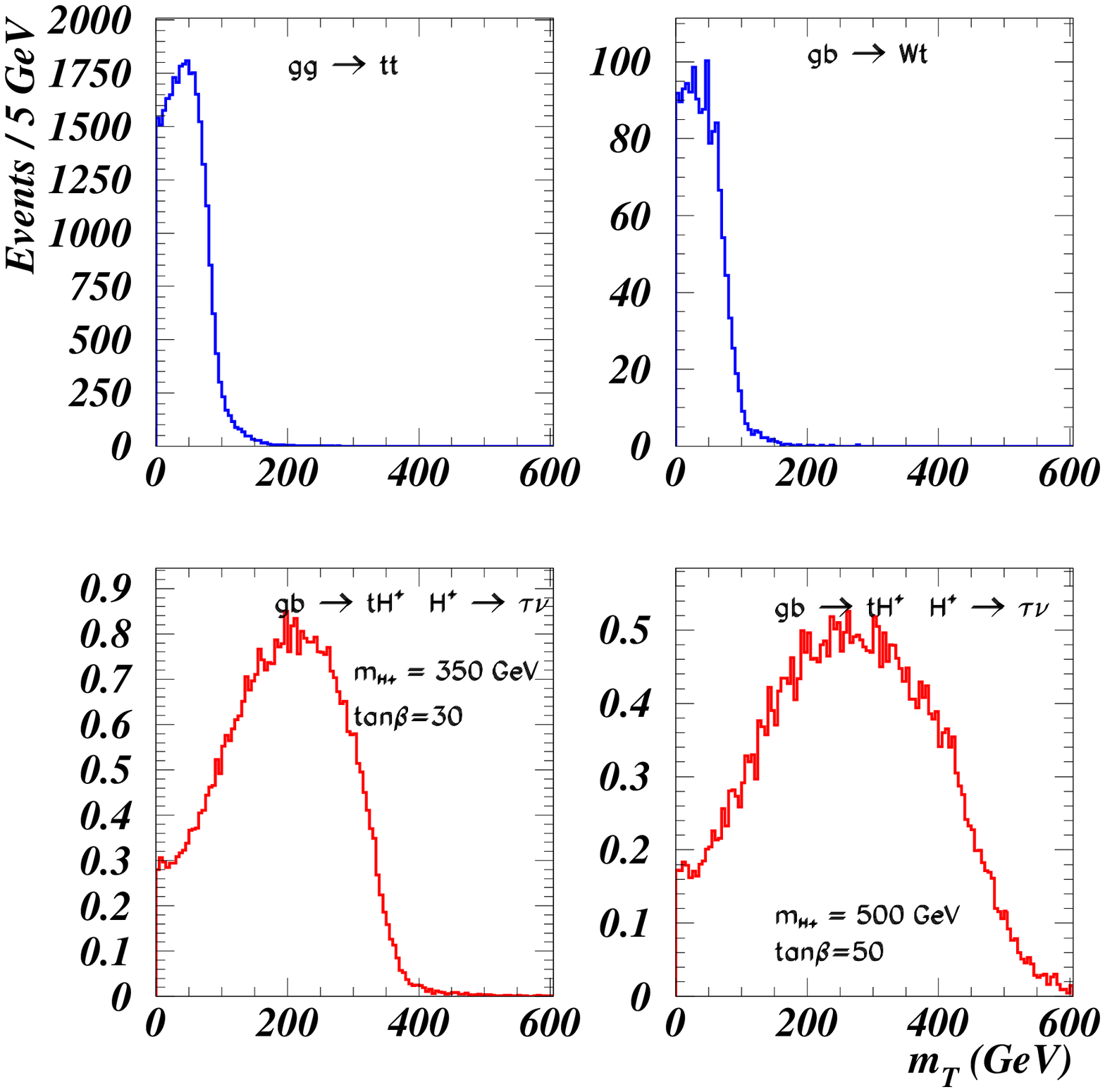}}&
\resizebox{6.3cm}{!}{\includegraphics{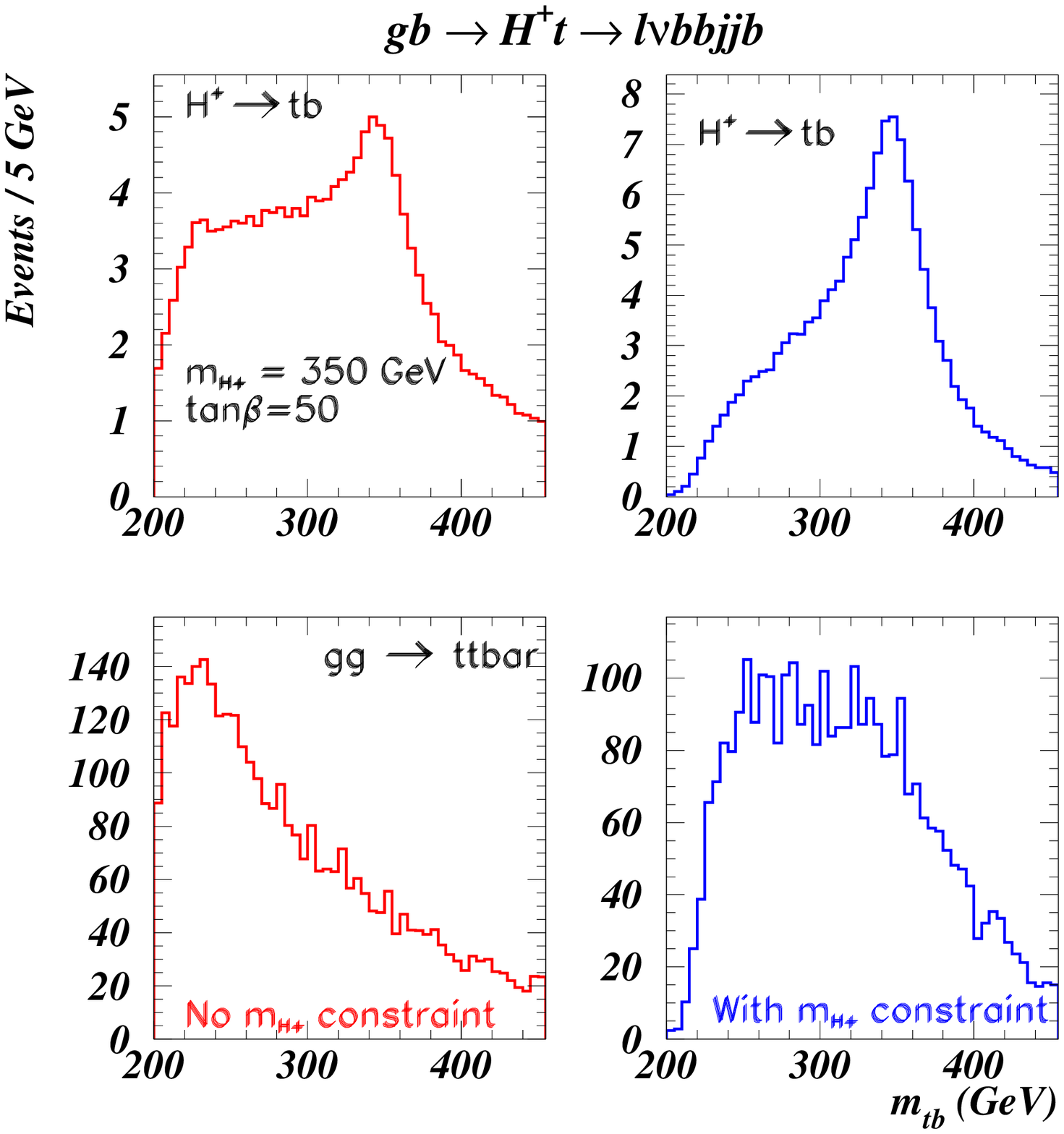}}\vspace*{-0.3cm}\\
{\small{(a)}} & {\small{(b)}}
\end{tabular}
\end{center}\vspace*{-1cm}
 \caption{a) Transverse mass $m_T$ distribution for signal and
   total background in the leptonic channel, taking into account the
   polarization of 
   the $\tau$-lepton, for an integrated luminosity of $30 \fb^{-1}$. 
b) The hadronic channel signal and the background distributions for the
   reconstructed invariant mass $m_{tb}$ of $\mHc=350$ GeV, 
 $\tan \beta =50$ and $30 \fb ^{-1}$.}
\label{fig:exp}
\end{figure}

The leptonic decay channel~(\ref{eq:leptonic}) provides the
best probe for the detection of such a state at the
LHC. There is a good signal reconstruction and background-free
environment for this channel. 
The main background processes in this channel are $gg\to
t\bar{t}\to jj\,b\, \tau\nu\bar{b}$ and 
$g\bar{b}\to W^+\bar{t}\to \tau^+\nu \bar{t}$. We have used 
the following trigger conditions: one hadronic $\tau$-jet; 
a $b$-tagged jet and at least two light jets.  
Further, we have the cut $\Delta\phi({p\!\!\!/}_T,p_T^{\tau}) > 1$ 
for background suppression, $\Delta\phi$ being 
the azimuthal opening angle between the $\tau$-jet 
and the missing transverse momentum, ${p\!\!\!/}_T$. 
The missing transverse
momentum is harder for the signal than for the background. These 
effects are well cumulated in the transverse mass, 
$m_T = \sqrt{2p_T^\tau{p\!\!\!/}_T\left[1-\cos(\Delta\phi)\right]}$,
which provides good discrimination between the signal and the 
backgrounds, as shown in Fig.~\ref{fig:exp}a. A final 
cut $m_T > 200$ GeV was used for the calculations of the signal-to-background
ratios and signal significances.

The production rates $\sigma\times
BR(H^+\to \tau^+\nu)\times BR(W\to jj)$ are shown in Table~\ref{tabletau}. 
We can see that the cuts have a high efficiency. 
In fact, signal rates after cuts are large enough to indeed consider  
$H^{+} \rightarrow \tau \nu$ a {\it golden channel} for the $H^{+}$
discovery at large $\tan \beta$. Despite the small branching
ratio $BR(H^+\to\tau^+\nu)$, the $\tau$-lepton affords an efficient
trigger to observe this channel. Subsequently, we assume that the
$H^{+}$ is discovered in this channel in order to reduce the
combinatorial background in the hadronic channel. 
\begin{table}[tb]\small
\begin{center}
\begin{tabular}{||l|c|c||c|c||}
\hline
& $\mHc=350$ GeV & $\mHc=500$ GeV & $t\bar{t}$ & $W^\pm t$
\\\hline
$\sigma\times BR$ & $99.9\fb$ & $30.7\fb$ & $79.1\pb$ & $16.3\pb$
\\\hline
Events  & 29958 & 9219  & $2.3 \times 10^{6}$ & $4.89 \times
10^{5}$ \\\hline
Events after cuts & 174 & 96 & 17 & 3 \\\hline
Efficiency & 0.6\%& 1\% & $8\times 10^{-6}$ & $6\times10^{-6}$ \\ \hline
$S/B$ & 7.9 & 4.4  \\\cline{1-3}
$S/\sqrt{B}$ & 37.1 & 20.5  \\ \cline{1-3}
Poisson & 23.1 & 14.6 \\\cline{1-3}
\end{tabular}
\end{center}\vspace*{-0.7cm}
\caption{The signal and background cross-sections, the number of events
  before cuts, the number of events after all cuts, the total efficiency, 
  the signal-to-background ratios ($S/B$), and the signal significances 
(Gaussian and Poisson) for the detection of the charged Higgs in the 
$\tau\nu$ channel at the LHC, 
for $300\fb^{-1}$ integrated luminosity and $\tb=50$.}
\label{tabletau}
\end{table}

\begin{table}[tb]
\small\begin{center}\vspace*{-0.2cm}
\begin{tabular}{||l|c|c||c||}
\hline
& $\mHc=350$ \,GeV & $\mHc=500$ \,GeV & $t\bar{t}q$ \\\hline
$\sigma\times BR$ & $248.4\fb$ &  $88\fb$ & $85\pb$
\\\hline
Events  & 74510 &  26389 & $2.55\times 10^{7}$ \\\hline
Events after cuts & 2100 & 784 &  59688
 \\\hline
Efficiency & 2.8\% & 3\% & 0.2\%
 \\ \hline
$S/B$ & 0.035 & 0.013   \\\cline{1-3}
 $S/\sqrt{B}$ & 8.6 & 3.2  \\ \cline{1-3}
\end{tabular}
\end{center}\vspace*{-0.6cm}
\caption{The signal and background cross-sections, the number of events 
before and after cuts, the total efficiency, $S/B$, and the signal 
  significances for the detection of the charged Higgs in the $tb$ channel at
  the LHC, for $300\fb^{-1}$ integrated luminosity and $\tb=50$.
\label{tabletb}}
\end{table}

The decay mode
$H^\pm \to t b$ has large QCD backgrounds that come
from $t\bar{t}q$ production with 
$t\bar{t}\rightarrow  Wb\,Wb\rightarrow l\nu b \,jj b$.  
We search for an isolated lepton, three $b$-tagged jets and 
at least two non-$b$-jets. The jet--jet combinations whose 
invariant masses are consistent with the $W$-boson mass, 
$|m_W-m_{jj}| < 25$ GeV are retained, and then 
we use the $W$-boson mass constraint to find the 
longitudinal component of the neutrino momentum in $W^\pm\to l\nu$. 
We keep the best two top-quark candidate that minimize 
$\chi^2 \equiv (m_t-m_{l\nu b})^2 + (m_t - m_{jjb})^2$.  
The remaining $b$-jet can be paired with either top quark to give 
two charged Higgs candidates, one of which leads to a combinatorial 
background.

Assuming that the charged Higgs is discovered through the $H^\pm \to \tau\nu$
channel and its mass determined from the $\tau\nu$ transverse mass 
distribution, the correct $H^+$ candidate in the $tb$ 
channel can be selected by using the measured $\mHc$ as a constraint. This is 
done by selecting the candidate whose invariant mass is closest to the 
measured charged Higgs mass: $\chi^2 =(m_{tb}-\mHc)^2$.
Fig.~\ref{fig:exp}b shows the signal distribution 
for the reconstructed invariant mass $m_{tb}$ for $\mHc=350$ GeV, 
$\tan \beta=50$. We can see on the right plots that $\mHc$ can be used
as a constraint to reduce the combinatorial background.
However,  some 
irreducible combinatorial noise still appears. In addition the $\mHc$
constraint reshapes the background distribution in 
$g g \rightarrow t \bar t X$ in such a way that no improvement in the 
signal-to-background ratio and signal significance is 
further observed. Therefore,
the knowledge of the shape and the normalization of the reshaped background 
would be necessary for the signal extraction. 
For these reasons, we did not use 
the $\mHc$ constraint for the results shown in this {work}. 

The subtraction of the background can then
be done by fitting the side bands and extrapolating in the 
signal region, which will be known from the $\mHc$ determination in the 
$H^\pm\to\tau\nu$ channel. However, 
this would be possible only for Higgs masses above $300$ GeV -- see 
Fig.~\ref{fig:exp}b. The signal and background results are 
summarized in Table~\ref{tabletb}. Even using  the $\mHc$ constraint, it is
difficult to observe $H^{\pm}$ signals in this channel above $\sim 400$
GeV. For masses above $\mHc\sim 400$ GeV the signal  
significance
can be enhanced by using the kinematics of the three-body production 
  process $gg\to H^+\bar{t}b$ \cite{jaumeroy}. 

Recent studies on the discovery prospects of a heavy charged MSSM Higgs
boson in the $H^+ \rightarrow \bar{t} b$ channel using three $b$-quarks  
tagging shows that, taking into account influences of systematic background
uncertainty, no discovery region is left in the MSSM parameter
space~\cite{steven}.  The (relatively) small $S/B$ and significances in
Table~\ref{tabletb} also seem to suggest it. However, for the 
$gg\to H^+\bar{t}b$ production 
process at the LHC, which together with the hadronic decay channel leads
to four $b$-tagged jets, no significant improvement in the discovery
potential with respect  to the three $b$-quarks in the final states has been
found~\cite{Nils}. Furthermore, in the $4$-$b$ tag case the $H^{+} \rightarrow
\tau \nu$ channel becomes less visible. Our strategy consists in
discovering (and measuring) the $H^\pm$ in the leptonic channel, and
afterwards measuring it in the hadronic channel. In this case the
background can be measured in the side bands with sufficient accuracy
(see above) and the hadronic channel is visible. Therefore, $3$-$b$ quarks 
in the final states is more successful in the analysis we are interested
in. 

The uncertainty in the ratio $R$ is dominated by the
reduced knowledge of the background shape and rate in the 
$H^{+} \rightarrow t b$ channel. We assume a theoretical uncertainty 
of $5\%$ on the branching ratios, $BR\,$s. Previous 
ATLAS studies have shown that the residual $gg \to t\bar{t}$ shape 
and normalization can be determined to $5\%$, and the scale
uncertainties on jet and lepton energies are 
expected to be of the order of $1 \%$ and $0.1 \%$
respectively~\cite{ketevi}. As explained above, for $m_{H^\pm} >
300$ GeV, the side band procedure can be used to subtract the 
residual background under the $H^{+}\rightarrow t b$ signal: we
assume also a $5\%$ uncertainty in the background subtraction method. Thus, the
statistical uncertainties can be estimated as $\sqrt{1/S}$. 
The cumulative results for the two channels
are summarized in Table~\ref{tablefinal} at an integrated
luminosity of $300 \fb^{-1}$.  Here,
the final result for the ratio $R$ is
  obtained by correcting the visible production rates after cuts for the
  total detection efficiency in Tables~\ref{tabletau} and~\ref{tabletb}
  and by the decay $BR\,$s of the $W$-bosons.
The simulation shows that the above-mentioned 
ratio can be measured with
an accuracy of $\sim 12$--$14 \%$ for $\tan \beta=50$, 
$\mHc=350$--$500$ GeV and at an integrated luminosity of $300 \fb^{-1}$. 

\begin{table}[tb]\small\begin{center}
\begin{tabular}{|l|c|c|}
\hline
& $\mHc=350$ \,GeV & $\mHc=500$ \,GeV \\\hline
Signals $\tau\nu/tb$ & 174 / 2100 = 0.08 &
96 / 784 = 0.12 \\ \hline
Signals (corrected) $\tau\nu/tb$ &  0.18 & 0.16 \\\hline
Systematic uncertainty & $\sim 9\%$ & $\sim 9\%$ \\ \hline
Total uncertainty & 12\% & 14\% \\\hline
Theory & 0.18 & 0.16 \\\hline
\end{tabular}
\end{center}\vspace*{-0.8cm}
\caption{Determination of the ratio~(\ref{eq:relation})
   for $300\fb^{-1}$ and $\tb=50$. Shown are: signal after cuts;
   signal after correcting for efficiencies and branching ratios;
   systematic uncertainty; total combined uncertainty; and
   theoretical prediction without SUSY corrections.\label{tablefinal}}
\end{table}

\begin{figure}[tb]
\centerline{\begin{tabular}{cc}
\resizebox{5cm}{!}{\includegraphics{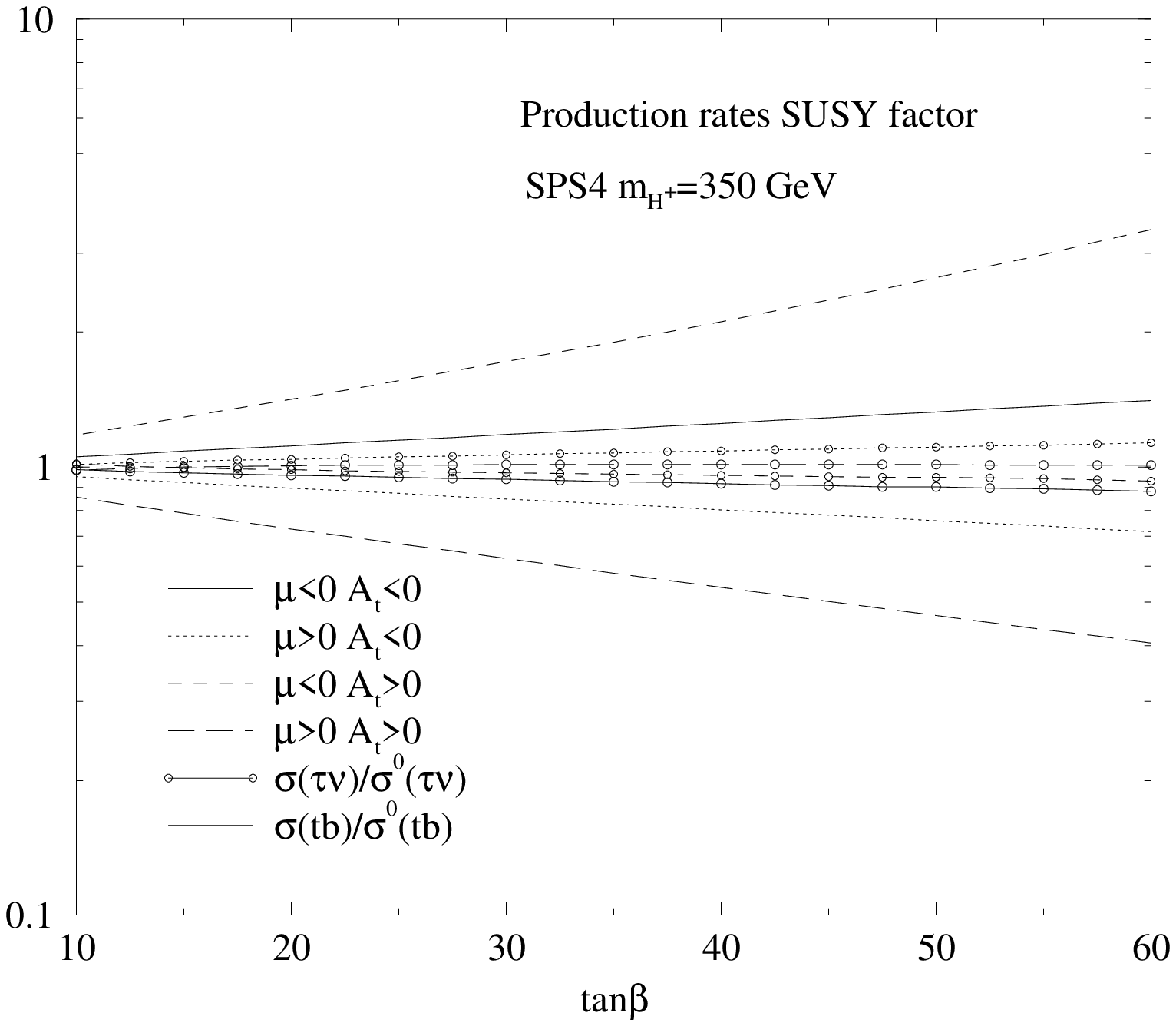}} &
\resizebox{5cm}{!}{\includegraphics{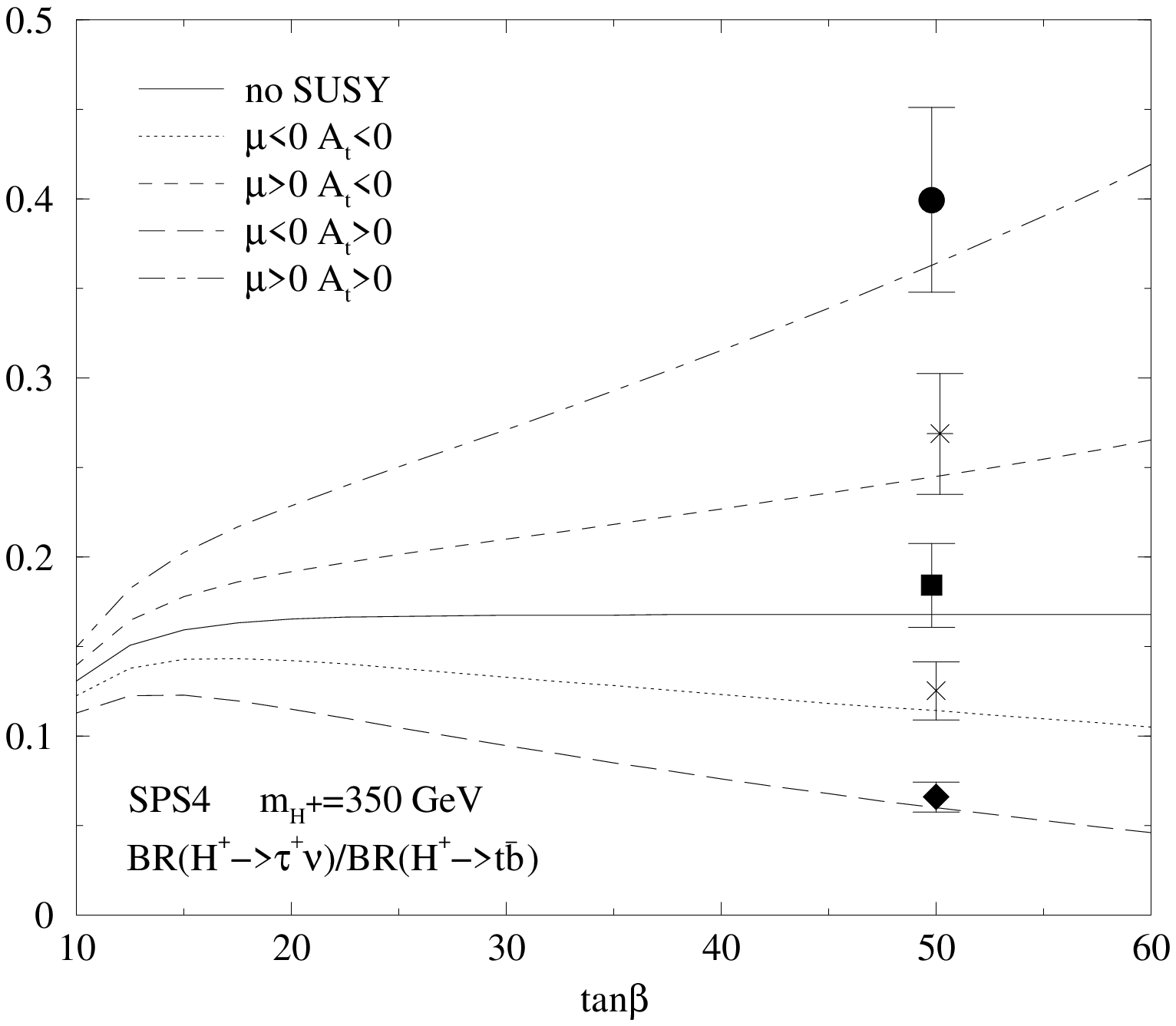}}\\
{\small{(a)}}&{\small{(b)}} 
\end{tabular}}\vspace*{-0.6cm}
\caption{a) Production rates enhancement/suppression factors
  for the $\tau$ and the $tb$ channels; b) the SUSY correction
  to the rate~(\ref{eq:relation2}). Plots as  functions of $\tb$ for
  $\mHc=350$ GeV and a SUSY spectrum as in SPS4, but for different
  choices of the signs of $\mu$ and $A_t$. Shown is also the
  experimental determination for each scenario.\label{fig:theory}}
\end{figure}

The role of the SUSY radiative corrections is twofold. Firstly,
by changing the
value of the Yukawa coupling, they change the value of the observable
$R$. Secondly, they change  
the value of the production cross-section $\sigma(pp\to H^+\bar{t}+X)$.
In Fig.~\ref{fig:theory}a, we show the 
enhancement/suppression factors as a
function of $\tb$ for $\mHc=350$ GeV and a SUSY mass spectrum defined as
SPS4 of the {\sl Snowmass Points and Slopes}
in \cite{Allanach:2002nj}, but choosing different scenarios for the
signs of $\mu$ and $A_t$.  
The production rate in the $\tau$-channel
 is fairly independent of the SUSY radiative corrections. 
On the contrary, the hadronic $t\bar{b}$ production channel 
receives large radiative
corrections. These corrections can be either positive (enhancing the
significance in Table~\ref{tabletb}) or negative 
(reducing it). This might permit to overcome the low signal-to-background
ratio of this channel in some SUSY scenarios.

Fig.~\ref{fig:theory}b shows the prediction for the ratio
$R$ as a function of $\tb$ for $\mHc=350$ GeV. 
We also show the experimental
determination carried out as before and repeated for each SUSY setup.
From Fig.~\ref{fig:theory}b it is clear that radiative SUSY
effects are visible at the LHC at a large significance. 
In particular, the $\mu<0$ scenarios can easily be discriminated, while the
  $\mu>0$ ones will be more difficult to establish, since the 
  signal rate of the hadronic channel is lower.
This feature then also allows for a measurement of the sign of the $\mu$
parameter, which can be determined by a $\sim 14 \%$ 
measurement of this ratio for $\tan \beta>30$. 

To summarize, we have quantitatively
shown that an LHC measurement of $R$ can give 
clear evidence for or against the SUSY nature of charged Higgs bosons. 

\bigskip 

\noindent{\small 
S.P. thanks the European Union for financial support (contract
MEIF-CT-2003-500030).}

\providecommand{\href}[2]{#2}

\end{document}